\documentclass[a4paper,11pt]{article}
\usepackage{pos}

\title{Benchmark of medium-induced radiative energy loss model for heavy-ion collisions}

\author*[a]{Iurii Karpenko}
\author[b]{Joerg Aichelin}
\author[b]{Pol Bernard Gossiaux}
\author[c]{Martin Rohrmoser}

\affiliation[a]{Faculty of Nuclear Sciences and Physical Engineering, Czech Technical University in Prague,\\  B\v rehov\'a 7, 11519 Prague 1, Czech Republic}

\affiliation[b]{SUBATECH, Universit\'e de Nantes, IMT Atlantique, IN2P3/CNRS,\\
4 rue Alfred Kastler, 44307 Nantes cedex 3, France}

\affiliation[c]{H.~Niewodnicza\'nski Institute of Nuclear Physics PAN, 31-342 Cracow, Poland}

\emailAdd{yu.karpenko@gmail.com}

\abstract{We report on a benchmark calculation of the in-medium radiative energy loss of low-virtuality jet partons within the EPOS3-Jet framework. The radiative energy loss is based on an extension of the Gunion-Bertsch matrix element for a massive projectile and a massive radiated gluon. On top of that, the coherence (LPM effect) is implemented by assigning a formation phase to the trial radiated gluons in a fashion similar to \cite{Zapp:2011ya} by Zapp, Stachel and Wiedemann. In a calculation with a simplified radiation kernel, we reproduce the radiation spectrum reported in \cite{Zapp:2011ya}. The radiation spectrum produces the LPM behaviour $dI/d\omega\propto\omega^{-1/2}$ up to an energy $\omega=\omega_c$, when the formation length of radiated gluons becomes comparable to the size of the medium. Beyond $\omega_c$, the radiation spectrum shows a characteristic suppression due to a smaller probability for a gluon to be formed in-medium.

Next, we embed the radiative energy loss of low-virtuality jet partons into a more realistic ``parton gun'' calculation, where a stream of hard partons at high initial energy $E_\text{ini}=100$~GeV and initial virtuality $Q^2=E^2$ passes through a box of QGP medium with a constant temperature. At the end of the box evolution, the partons are hadronized using Pythia 8, and the jets are reconstructed with the FASTJET package. We find that the full jet energy loss in such scenario approaches a ballpark value reported by the ALICE collaboration. However, the calculation uses a somewhat larger value of the coupling constant $\alpha_s$ to compensate for the missing collisional energy loss of the low-virtuality jet partons.}

\FullConference{%
  HardProbes2020\\
  1-5 June 2020\\
  Austin, Texas}

\newcommand{\vkt}{\vec{k_T}}
\newcommand{\vlt}{\vec{l_T}}

\begin{document}
\maketitle

{\bf Introduction:} In the last decade, both experimental extraction and theoretical modeling of jets in heavy-ion collisions became more and more refined. At present, one has understood that the description of the fine substructure properties of the jets, such as the jet shape, requires a rather precise description of both, the underlying space-time evolution of the medium and the back reaction of the jet on the medium, sometimes called  ``medium recoil''. Our aim is to create a comprehensive model of jet and medium evolution, and in these proceedings we report on the development of one of its core components: the process of medium-induced gluon radiation by low-virtuality jet partons.

{\bf The model:} The medium-induced radiation is based on  the emission of gluons by massive quarks, which move through and interact with light medium quarks and medium gluons \cite{Aichelin:2013mra}. The latter extends the known calculation by Gunion and Bertsch \cite{Gunion:1981qs} to the case of a massive projectile and a massive radiated gluon.

A direct computation of the matrix elements corresponding to the gluon bremsstrahlung is difficult. A simplifying assumption used in \cite{Aichelin:2013mra} is that, in the region of small $x$, the matrix elements from QCD can be approximated by a so-called \textit{scalar} QCD. The scalar QCD (SQCD) is a case of spin-0 quarks interacting with a non-Abelian gauge field (gluons). At high enough energy, the SQCD leads to a factorized formula for the total cross section of the radiation process:
\begin{equation} \frac{d\sigma^{Qq\rightarrow Qqg}}{dx d^2k_T d^2l_T}=\frac{d\sigma_\text{el}}{d^2 l_T}P_g(x,k_T,l_T)\theta(\Delta), \label{eq:dSigma}
\end{equation}
where $k_T$ is the transverse momentum of the radiated gluon, $l_T$ is the transverse momentum acquired by the medium parton, $\frac{d\sigma_{\rm el}}{d^2 l_T}\rightarrow\frac{8\alpha_s^2}{9(\vlt^2+\mu^2)^2}$ and
\begin{equation}
\Delta=\left(x(1-x)\,s-x\,M^2-\vkt^2+2x\,\vkt\cdot\vlt\right)^2
-4x(1-x)\,\vlt^2\,(x\,s-\vkt^2).
\label{eq:def_Delta}
\end{equation}
The $\Theta $ function encodes the kinematic limits.
The part corresponding to the gluon radiation is:
\begin{equation}
P_g(x,\vkt,\vlt;M)=\frac{C_A\alpha_s}{\pi^2}\frac{1-x}{x}
\left(\frac{\vkt}{\vkt^2+x^2 M^2}
-\frac{\vkt-\vlt}{(\vkt-\vlt)^2 +x^2 M^2}\right)^2,
\label{eq:gluon_distribution}
\end{equation}
where $M$ is the mass of the projectile $Q$. An important feature of the expressions above is that they can be easily extended for the case of a massive radiated gluon by substituting $x^2M^2\rightarrow x^2M^2+(1-x)m_g^2$, where $m_g$ is the gluon mass. As one can see, the mass of the projectile quark and the gluon enters the expressions explicitly. However, in the present proceedings we do not study the hierarchy of the energy loss by projectiles with different masses.

On top of the radiation kernel described by the formulas above, we have implemented the coherent gluon radiation via multiple scatterings of the projectile with the medium partons. More details and tests of the procedure will be reported in an upcoming publication, whereas here we list the main details:
\begin{itemize}
 \item The formation and evolution of the parton shower from the initial hard parton proceeds in time-steps. Such feature makes it very convenient to couple the parton shower to a realistic hydrodynamic evolution of the medium, which is typically performed numerically. 
 \item At each time-step, we decide whether an elastic scattering or an incoherent inelastic scattering takes place, according to probabilities $\Gamma_{\rm el} \Delta t$ and $\Gamma_{\rm inel} \Delta t$, respectively. The result of such incoherent inelastic scattering is a trial radiated (or unformed) gluon.
 \item Each of the trial radiated gluons starts its evolution with an initial phase $\varphi=0$ and a collision counter $N_s=1$. The projectile may have more than one trial radiated gluon associated with it, and the gluons are treated independently and in parallel.
 \item Each of the trial radiated gluons also experiences elastic scatterings with the medium partons, according to the same elastic rate $\Gamma_{\rm coll}$. Each scattering changes the transverse momentum $k_T$ of the trial radiated gluon, and increments its the collision counter $N_s$ by one.  The phase of a trial radiated gluon increases by $\Delta\varphi=\omega/k_T^2\cdot\Delta t$.
 \item Once a trial radiated gluon accumulates formation phase $\varphi=3.0$ while still being inside the medium, it is considered to be a candidate for a real radiated gluon. At this moment, the trial radiated gluon is either accepted as a real radiated gluon with a probability $1/N_s$, or is discarded with a probability $1-1/N_s$.
 \item If the trial radiated gluon is accepted, its initial energy-momentum are subtracted from the energy-momentum of the projectile.
\end{itemize}
Such an algorithm allows to mimic the process of coherent radiation in a probabilistic fashion \cite{Zapp:2011ya}. To test the algorithm, we first set it up to reproduce an exact result for the energy spectrum of radiated gluons in the BDMPS-Z limit. To do so, we reduce the radiation kernel to $dN/(d\omega dk_T)=1/\omega$, $k_T=0$ and set the total incoherent radiation rate to be $\Gamma_\text{el}=\Gamma_\text{inel}=0.1$~fm. The initial transverse momentum of gluons is set to zero, $k_T=0$, in accordance with \cite{Zapp:2011ya} and assuming that the transverse momentum is picked up mostly by the multiple elastic scatterings with the medium partons.

\begin{figure}
\includegraphics[width=1.0\textwidth]{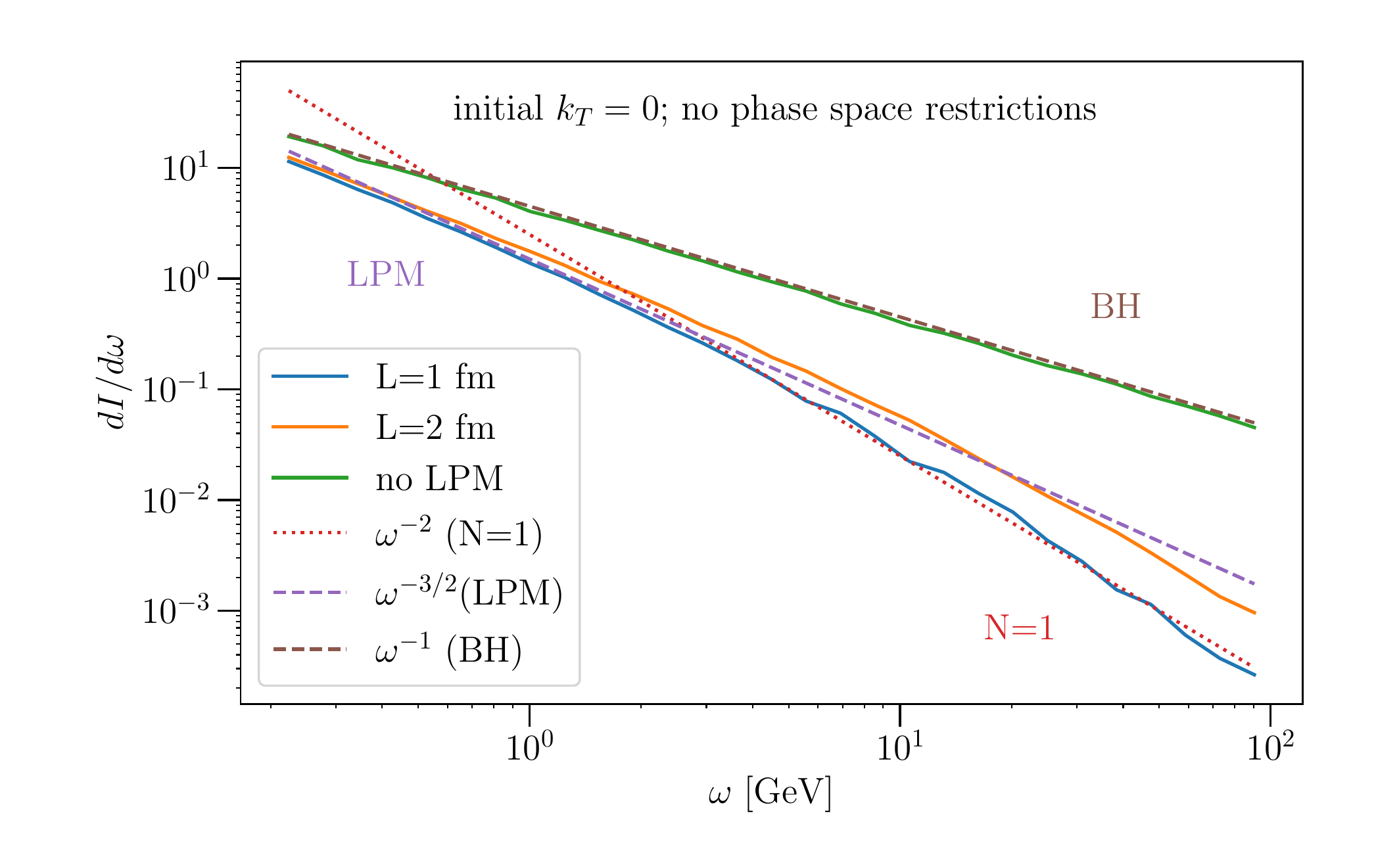}\vspace{-10pt}
\caption{Energy spectrum of gluons, radiated by a low-virtuality projectile with initial energy $E_\text{ini}=100$~GeV in a medium with lengths $L=1,2$~fm and a fixed inelastic mean free path $\Gamma_\text{inel}=0.1$~fm.}\label{fig1}
\end{figure}

{\bf Results and discussion:} With the setup above and the $\varphi=0.3$ setting, we run a so-called ``parton gun'' simulation, shooting mono-energetic on-shell quarks with $E=100$~GeV into a box of medium, where the temperature is adjusted so that the $\hat{q}=0.4$~GeV$^2$/fm. The resulting intensity spectra of the radiated gluons, as a function of the gluon energy $\omega$, are plotted in Fig.~\ref{fig1}. A transition from the coherent regime (dubbed as ``LPM'' in the plot) to the regime where at most 1 gluon is being emitted (dubbed as ``N=1'' in the plot), happens around gluon's energy:
\[ \omega_c\approx \frac{\hat q L^2}{2\varphi_f \hbar}. \]
With the present settings and a box size $L=1$~fm,  $\omega_c\approx3.4$~GeV, whereas $L=2$~fm corresponds to $\omega_c\approx13.5$~GeV. One can clearly see on the plot, that indeed in the  $L=1$~fm case the transition happens around the $\omega_c$ value, which was calculated above.  Also, by setting the formation phase condition $\varphi=0$, we can restore the incoherent radiation regime (which is a green curve on the plot), where the radiation spectrum follows the known $1/\omega$ dependence.

Next, we turn to a more realistic scenario, where the jet evolution starts from an energetic parton with $E=100$~GeV and initial virtuality scale $Q_\text{ini}=E$. As such, the process of jet formation - subsequent parton splittings via a virtuality-ordered parton shower - takes place before the partons at the lowest virtuality scale, $Q_0=0.3$~GeV, start to radiate gluons due to medium interactions. In this scenario, the medium is modelled as a gas of $N_f=3$ massive quarks with $m_q(T)=330$~MeV and $N_c=3$ gluons with mass $m_g(T)=564$~MeV, at a temperature of $T=350$~MeV. The size of the medium is taken as $L=4$~fm, which very roughly mimics the average mean free path of the energetic partons in the dense QGP medium.

In the realistic case, the medium-induced gluon radiation is calculated according to Eqs.~\ref{eq:dSigma}-\ref{eq:gluon_distribution} with the infrared regulator $\mu=623$~MeV (which is proportional to the Debye mass). We take a somewhat larger value of the coupling constant with respect to other approaches, $\alpha_s=0.4$, which is partially justified by the fact that for this calculation, the jet partons lose energy only via the medium-induced radiation (whereas  only the radiated trial gluons suffer elastic scatterings). Coherence effects are also included as described above, with the value of the formation phase $\varphi=3$. At the end of the in-medium evolution, the jet partons are hadronized with Pythia 8 \cite{Sjostrand:2014zea}. Finally, the jets are reconstructed from the final-state jet hadrons using the anti-$k_T$ algorithm with a cone size $R=0.4$ via the FASTJET package \cite{Cacciari:2011ma}.

\begin{figure}
\begin{center}
\includegraphics[width=0.8\textwidth]{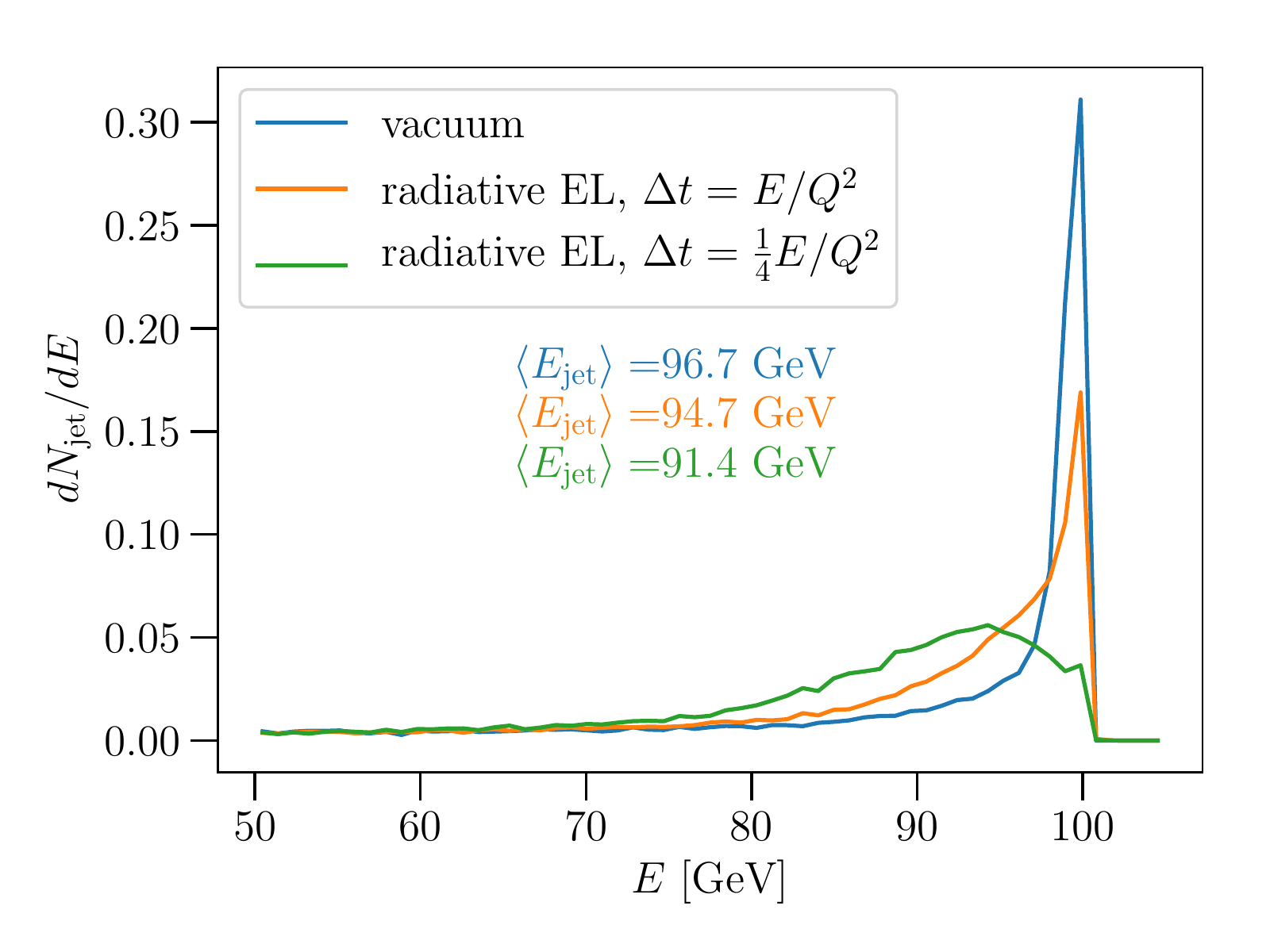}\vspace{-10pt}
\end{center}
\caption{Profile (distribution) of radiative energy loss by a projectile with initial energy $E_\text{ini}=100$~GeV in a QGP medium with length $L=4$~fm and a temperature $T=350$~MeV.}\label{fig2}
\end{figure}

We plot the energy distribution of the reconstructed jets - both with and without medium-induced energy loss - in Fig.~\ref{fig2}. On this plot, we also show a third scenario (represented by the green curve), where the formation time of the jet partons is taken to be 4 times smaller than the standard prescription $\Delta t=E/Q^2$. First we note that the profiles of the jet energy in both,the  vacuum and the ``standard'' medium-modified case,  increase monotonically with the energy of the reconstructed jet. In the scenario of the short jet formation (the green curve), there is a broad peak around $E\approx94$~GeV. Such profile of the radiative energy loss of a jet is qualitatively consistent with an eariler studies in AMY formalism with a similar setup \cite{Qin:2007rn}. We note that the difference between the mean energy of the reconstructed ``vacuum'' jet and the reconstructed medium-modified jet with $\Delta t=\frac{1}{4}E/Q^2$ is in the same ballpark as the $p_T$ shift of the jet spectrum, measured in Pb-Pb collisions in the jet $p_T$ range [60,100] GeV by ALICE \cite{Adam:2015doa}.

{\bf Acknowledgements:} We acknowledge support by the project Centre of Advanced Applied Sciences with number CZ.02.1.01/0.0/0.0/16-019/0000778, which is co-financed by the European Union, and from the European Union’s Horizon 2020 research and innovation program STRONG-2020 under grant agreement No 824093.

\begin{thebibliography}{99}
\bibitem{Aichelin:2013mra}
J.~Aichelin, P.~B.~Gossiaux and T.~Gousset,
Phys. Rev. D \textbf{89} (2014) no.7, 074018

\bibitem{Gunion:1981qs}
J.~F.~Gunion and G.~Bertsch,
Phys. Rev. D \textbf{25} (1982), 746

\bibitem{Zapp:2011ya}
K.~C.~Zapp, J.~Stachel and U.~A.~Wiedemann,
JHEP \textbf{07} (2011), 118

\bibitem{Sjostrand:2014zea}
T.~Sjöstrand, S.~Ask, J.~R.~Christiansen, R.~Corke, N.~Desai, P.~Ilten, S.~Mrenna, S.~Prestel, C.~O.~Rasmussen and P.~Z.~Skands,
Comput. Phys. Commun. \textbf{191} (2015), 159-177

\bibitem{Cacciari:2011ma}
M.~Cacciari, G.~P.~Salam and G.~Soyez,
Eur. Phys. J. C \textbf{72} (2012), 1896

\bibitem{Qin:2007rn}
G.~Y.~Qin, J.~Ruppert, C.~Gale, S.~Jeon, G.~D.~Moore and M.~G.~Mustafa,
Phys. Rev. Lett. \textbf{100} (2008), 072301

\bibitem{Adam:2015doa}
J.~Adam \textit{et al.} [ALICE],
JHEP \textbf{09} (2015), 170
doi:10.1007/JHEP09(2015)170
\end{thebibliography}
\end{document}